\documentclass[12pt]{article}

\catcode`@=11
\newcount\@tempcntc
\def\@citex[#1]#2{\if@filesw\immediate\write\@auxout{\string\citation{#2}}\fi
  \@tempcnta\z@\@tempcntb\m@ne\def\@citea{}\@cite{\@for\@citeb:=#2\do
    {\@ifundefined
       {b@\@citeb}{\@citeo\@tempcntb\m@ne\@citea\def\@citea{,}{\bf
?}\@warning
       {Citation `\@citeb' on page \thepage \space undefined}}%
    {\setbox\z@\hbox{\global\@tempcntc0\csname b@\@citeb\endcsname\relax}%
     \ifnum\@tempcntc=\z@ \@citeo\@tempcntb\m@ne
       \@citea\def\@citea{,}\hbox{\csname b@\@citeb\endcsname}%
     \else
      \advance\@tempcntb\@ne
      \ifnum\@tempcntb=\@tempcntc
      \else\advance\@tempcntb\m@ne\@citeo
      \@tempcnta\@tempcntc\@tempcntb\@tempcntc\fi\fi}}\@citeo}{#1}}
\def\@citeo{\ifnum\@tempcnta>\@tempcntb\else\@citea\def\@citea{,}%
  \ifnum\@tempcnta=\@tempcntb\the\@tempcnta\else
   {\advance\@tempcnta\@ne\ifnum\@tempcnta=\@tempcntb \else
\def\@citea{--}\fi
    \advance\@tempcnta\m@ne\the\@tempcnta\@citea\the\@tempcntb}\fi\fi}
\catcode`@=12

\begin{document}
\title{\vskip-3cm{\baselineskip12pt
\centerline{\normalsize DESY 00-074\hfill ISSN 0418-9833}
\centerline{\normalsize NYU/00/05/01\hfill}
\centerline{\normalsize hep-th/0005149\hfill}
\centerline{\normalsize May 2000\hfill}}
\vskip1.5cm 
Width and Partial Widths of Unstable Particles}
\author{{\sc Pietro A. Grassi,$^1$ Bernd A. Kniehl,$^2$ Alberto Sirlin$^1$}\\
{\normalsize $^1$ Department of Physics, New York University,}\\
{\normalsize 4~Washington Place, New York, NY~10003, USA}\\
{\normalsize $^2$ II. Institut f\"ur Theoretische Physik, Universit\"at
Hamburg,}\\
{\normalsize Luruper Chaussee 149, 22761 Hamburg, Germany}}

\date{}

\maketitle
\thispagestyle{empty}

\begin{abstract}
In the gauge theory context, a definition of branching ratios and partial
widths of unstable particles is proposed that satisfies the basic principles
of additivity and gauge independence.
A simpler definition, similar to the conventional one, is examined in the
$Z^0$-boson case. In order to establish contact with experiment, we show that 
it leads to a peak cross section that justifies the expression used by the
LEP Electroweak Working Group through next-to-next-to-leading order, provided
that the {\it pole} rather than the {\it on-shell} mass and width of the $Z^0$
boson are employed.

\medskip

\end{abstract}

\newpage

The mass, width, and partial widths of unstable particles rank among the basic
concepts in particle physics.
In fact, most fundamental particles of nature are unstable, and their masses,
widths, and partial widths are some of their crucial defining properties. 
Yet, the precise and consistent definitions of these concepts have been
notoriously difficult and elusive over a period spanning several decades.
The reason is that unstable particles are not asymptotic states and,
consequently, they lie somewhat outside the traditional formulation of Quantum
Field Theory.

The conventional definitions of mass and width are
\begin{eqnarray}
\label{def_mass_1}
M^2&=&M^2_0+{\rm Re}\, A(M^2),\\
\label{def_width_1}
M\Gamma&=&-\frac{{\rm Im}\, A(M^2)}{1-{\rm Re}\, A'(M^2)},  
\end{eqnarray}
where $M_0$ is the bare mass and $A(s)$ is the self-energy in the case of
scalar bosons and the transverse self-energy in the case of vector bosons.
The partial widths are then defined by decomposing the numerator of
Eq.~(\ref{def_width_1}) into a sum of contributions involving distinct sets of
final-state physical particles.
Most calculations of partial and total widths are based on
Eqs.~(\ref{def_mass_1}) and (\ref{def_width_1}).
We will refer to $M$ as the on-shell mass and to Eqs.~(\ref{def_mass_1}) and
(\ref{def_width_1}) as the conventional on-shell formulation.

The emergence of gauge theories has brought into the discussion a new and
powerful element, namely the requirement of gauge independence of physical
observables.
It was shown in Ref.~\cite{si_1} that, in a gauge theory,
Eqs.~(\ref{def_mass_1}) and (\ref{def_width_1}) become gauge dependent in
$O(g^4)$ and $O(g^6)$, respectively, where $g$ is a generic gauge coupling.
As the leading contributions to $M^2$ and $\Gamma$ are of order $O(g^0)$ and
$O(g^2)$, respectively, we see that in both cases the problem arises in the
next-to-next-to-leading order (NNLO).
In the same paper, it was proposed that the way of solving this predicament is
to base the definitions of mass and width on the complex-valued position of
the propagator's pole,
\begin{equation}
\label{def_mass_2}
\bar s=M^2_0+A(\bar s), 
\end{equation}
an idea that goes back to well known tenets of $S$-matrix theory
\cite{s-matrix}.
A frequently employed parameterization is $\bar s=m_2^2-im_2\Gamma_2$, where
we use the notation of Ref.~\cite{si_1}.
Identifying $m_2$ and $\Gamma_2$ with the gauge-independent definitions of
mass and width of the unstable particle, it follows from
Eq.~(\ref{def_mass_2}) that
\begin{equation}
\label{def_width_2}
m_2\Gamma_2=-{\rm Im}\,A(\bar s).  
\end{equation}
Over the last several years, a number of authors have advocated the use of
$\bar s$ as the basis for the definition of mass and width \cite{zmass}, the
conclusions of Ref.~\cite{si_1} have been confirmed by later studies
\cite{psz,si-kniehl} and proven to all orders \cite{gg}.
It has been shown that, in the case of a heavy Higgs boson, the gauge
dependences of $M$ and $\Gamma$ are numerically large \cite{si-kniehl}.
It has also been emphasized that the on-shell definition of width
[Eq.~(\ref{def_width_1})] leads to severe problems if $A(s)$ is not analytic
in the neighborhood of $M^2$.
This occurs, for instance, when the mass of the decaying particle lies very
close to a physical threshold \cite{bhatta,pal} or, in the resonance region,
when the unstable particle is coupled to massless quanta, as in the cases of
the $W$ boson and the unstable quarks \cite{psw}.
Significant progress has also been achieved in the treatment of unstable
particles in the framework of the Pinch Technique \cite{pt}.

An important issue that arises at this stage is the following:
if Eq.~(\ref{def_width_2}) provides a consistent definition of width, what is
the definition of partial widths?
It must clearly satisfy two important properties: additivity, {\it i.e.}\ the
sum of the partial widths must equal the total width
[Eq.~(\ref{def_width_2})], and gauge independence.

We consider the process $i\to Z^0\to f$, where $i$ and $f$ are
initial and final states involving particles which are either stable or have 
negligible widths.
The transverse part of the propagator is given by 
\begin{equation}
\label{new_1}
{\cal D}_{\mu\nu}=-i\frac{Q_{\mu\nu}}{s-\bar s-[A(s)-A(\bar s)]}, 
\end{equation}
where $Q_{\mu\nu}=g_{\mu\nu}-p_\mu p_\nu/s$, $p_\mu$ is the four-momentum of
the $Z^0$ boson, and $s=p^2$.
The vertex amplitude is of the form 
\begin{equation}
\label{new_2}
V_f^\mu(s)\equiv\left\langle f\left|J^\mu_Z\right|0\right\rangle
=\sum_av^{(a)}_f(s,\dots)M^{(a)\mu}_f,
\end{equation}
where $M^{(a)\mu}_f$ denote various independent vector and axial-vector matrix
elements involving the spinors, polarization four-vectors, and four-momenta of
the final-state particles, while $v^{(a)}_f(s,\dots)$ are scalar functions.
The dots indicate their additional dependence on the momenta of the 
final-state particles. 
In this paper, we use the convention of including the coupling $-ig/c$, where
$c$ is an abbreviation for $\cos\theta_w$, in the definition of
$v^{(a)}_f(s,\dots)$, so that, in leading order, $v^{(a)}_f(s,\dots)=O(g)$.
Expanding Eq.~(\ref{new_2}) and the denominator of Eq.~(\ref{new_1}) about
$s=\bar s$, it is well-known \cite{zmass} that the overall amplitude can be written in the
form 
\begin{equation}\label{new_3}
{\cal A}_{fi}(s)=-i\frac{Q_{\mu\nu}V_f^\mu(\bar s)V_i^\nu(\bar s)}
{(s-\bar s)[1-A^\prime(\bar s)]}+N,
\end{equation}
where $N$ stands for non-resonant contributions. 
As the pole residues
$v^{(a)}_f(\bar s,\dots)v^{(b)}_i(\bar s,\dots)/[1-A^\prime(\bar s)]$ are
gauge independent for any choice of the states $i,f$ and the amplitudes $a,b$,
a gauge-independent definition of partial width is given by 
\begin{equation}
\label{new_4}
m_2\hat\Gamma_f=-{1\over 6}\sum_{\rm spins}\int d\Phi_f 
\frac{Q_{\mu\nu}V_f^{\mu*}(\bar s)V_f^\nu(\bar s)}
{\left|1-A^\prime(\bar s)\right|}.
\end{equation}
The integration is over the phase space of the final-state particles with
$(\sum_n p_n)^2=m^2_2$, a factor of $1/3$ arises from the average over the
initial-state polarization, and a factor of $1/2$ from the familiar relation
between $m_2\hat\Gamma_f$ and the integrated amplitude square. 

A limitation of Eq.~(\ref{new_4}) is that there is no guarantee that it
satisfies the additivity property.
In fact, it is expected that $\sum_f\hat\Gamma_f\neq\Gamma_2$ when one
includes NNLO contributions.
In order to remedy this situation, we propose to define the branching ratios
by 
\begin{equation}
\label{new_5}
B_f=\frac{\hat\Gamma_f}{\sum_f\hat\Gamma_f},
\end{equation}
and the partial widths by 
\begin{equation}
\label{new_6}
\Gamma_f=B_f\Gamma_2.  
\end{equation}
The gauge independence of Eq.~(\ref{new_4}) implies that of Eq.~(\ref{new_5}),
while Eqs.~(\ref{new_5}) and (\ref{new_6}) guarantee the additivity property,
$\sum_f\Gamma_f=\Gamma_2$.
The rescaling in Eqs.~(\ref{new_5}) and (\ref{new_6}) implies that
$\hat\Gamma_f=\Gamma_f(X/m_2\Gamma_2)$, where
\begin{equation}
\label{new_7}
X=m_2\sum_f\hat\Gamma_f.
\end{equation}

The resonant cross section at $s=m^2_2$ is proportional to 
$\hat\Gamma_e\hat\Gamma_f/(m_2\Gamma_2)^2 
=\left[\Gamma_e\Gamma_f/(m_2\Gamma_2)^2\right]$
$\times(X/m_2\Gamma_2)^2$, where
$\Gamma_e$ is the $Z^0\to e^+e^-$ partial width.
Hence, it is modified by a factor $(X/m_2\Gamma_2)^2$ when expressed in terms
of the widths $\Gamma_f$ that satisfy the additivity property.
We note that $X/m_2$ and $\Gamma_2$ represent two different definitions of
total width, based on the pole residues and the pole position, respectively.
The ratio $(X/m_2\Gamma_2)$ differs from unity by gauge-independent terms of
$O(g^4)$, {\it i.e.}\ in NNLO.
As a consequence, in the $Z^0$-boson case, this is expected to be a very small
effect, of the same order of magnitude as non-resonant contributions that are
frequently neglected.

Next, we examine the difference $X-m_2\Gamma_2$ in greater detail. 
It is convenient to split
$I\left(m^2_2\right)\equiv{\rm Im}\,A\left(m^2_2\right)$ in the form
\begin{eqnarray}
\label{new_9}
I\left(m^2_2\right)&=&F\left(m^2_2\right)+G\left(m^2_2\right),\\
F\left(m^2_2\right)&=&\sum_fI_f\left(m^2_2\right),\\
I_f\left(m^2_2\right)&=&\frac{1}{6}\sum_{\rm spins}\int d\Phi_f 
Q_{\mu\nu}V_f^{\mu*}\left(m_2^2\right)V_f^\nu\left(m_2^2\right). \label{new_11}
\end{eqnarray}
$-I_f(m^2_2)/m_2$ is the conventional expression for the partial width of the
unstable particle into the physical state $f$, modulo the wave-function
renormalization of the unstable particle, with the important difference that
it is evaluated at the gauge-independent pole mass $m_2$ rather than the
gauge-dependent on-shell mass $M$.
$G\left(m^2_2\right)\equiv I\left(m^2_2\right)-F\left(m^2_2\right)$ involves
contributions from unphysical intermediate states (Goldstone bosons,
Faddeev-Popov ghosts, and longitudinal modes of gauge bosons), which can
contribute to $I\left(m^2_2\right)$ for sufficiently low values of the gauge-parameters
\cite{si_1}.

In the conventional on-shell formulation, it is assumed that $I(M^2)$ can be
expressed as a sum of contributions involving solely physical intermediate
states, namely $I(M^2)=\sum_fI_f(M^2)$.
The argument invokes the unitarity of the $S$ matrix and would, in fact, be
valid if $I(M^2)$ were an $S$-matrix amplitude.
However, as the unstable particle is not an asymptotic state, this is not the
case, and the above decomposition into physical cut contributions must be
viewed as an approximation.
In fact, we will show in this section that $G\left(m^2_2\right)\neq0$ in
$O(g^6)$, {\it i.e.}\ in NNLO. 

We study the difference $X-m_2\Gamma_2=X+{\rm Im}\,A(\bar s)$, where $X$ is
defined in Eqs.~(\ref{new_4}) and (\ref{new_7}), by expanding
$v^{(a)}_f(\bar s,\dots)$, $A^\prime(\bar s)$, and $A(\bar s)$ about $s=m^2_2$
through terms of $O(g^6)$.
The leading terms in the expansion of ${\rm Im}\,A(\bar s)$ is
$I\left(m^2_2\right)$, for which we employ the decomposition of
Eq.~(\ref{new_9}). 
The $F\left(m^2_2\right)$ term cancels the leading contribution from $X$, and
we find 
\begin{eqnarray}
\label{new_13}
&&X-m_2\Gamma_2=\frac{(m_2\Gamma_2)^2}{2}I^{\prime\prime}
-\frac{m_2\Gamma_2}{2}(I^\prime)^2+G\left(m^2_2\right)
\nonumber\\
&&{}-\frac{m_2\Gamma_2}{3}\sum_{f,\rm spins}\int d\Phi_fQ_{\mu\nu} 
{\rm Im}\,\left[V_f^{\mu*}\left(m_2^2\right)V_f^{\nu\prime}\left(m_2^2\right)
\right],
\end{eqnarray}
where the primes indicate derivatives with respect to $s$, evaluated at
$s=m^2_2$.
Since $X$ and $m_2\Gamma_2$ are gauge independent, Eq.~(\ref{new_13})
determines the gauge-dependent part of $G\left(m^2_2\right)$ in $O(g^6)$.
As $\Gamma_2$ and $I$ are of $O(g^2)$, $v^{(a)}_f$ is of $O(g)$, and
$v^{(a)\prime}_f$ is of $O(g^3)$, it suffices to consider the gauge dependence
of the one-loop electroweak contributions to $I(s)$ and $v^{(a)}_f(s,\dots)$. 
Furthermore, in the consideration of the vertex contributions, we may restrict
ourselves to the two-particle final states, since those involving more
particles give gauge-dependent contributions of higher order. 

In the $Z^0$-boson case, the gauge dependence of $I(s)$ and
$v^{(a)}_f(s,\dots)$ at the one-loop level can be obtained from
Eqs.~(7), (8), (17), and (24) of Ref.~\cite{deg} in the approximation of
neglecting the masses of the external fermions, which we henceforth adopt.
Applying those results to Eq.~(\ref{new_13}) and noting that
$G\left(m^2_2\right)$ vanishes for $\xi_W>1/c^2$, where $\xi_W$ is the gauge
parameter associated with the $W$ boson, we find
\begin{eqnarray}
G\left(m^2_2\right)
&=&\frac{m_2\Gamma_2}{2}\left[(I^\prime)^2-(F^\prime)^2\right]
-g^2c^2m^2_2\Gamma^2_2(\xi_W-1)
\nonumber\\
&&{}\times{\rm Im}\,\eta_W\left(m^2_2\right)+O(g^8),
\label{new_14}
\end{eqnarray}
where $\eta_W$ is a gauge-dependent amplitude given in Ref.~\cite{deg}.
Its imaginary part is non-vanishing in a subclass of gauges characterized by
$M_Z>2\sqrt{\xi_W}M_W$ or $\xi_W<1/(4c^2)$ \cite{si_1,deg}.
It is worth noting that one-loop $\gamma$--$Z$ mixing contributions, which
have been taken into account, cancel in the derivation.

We now discuss an alternative and manifestly additive definition of 
branching ratios, namely
\begin{equation}
\label{new_15}
\tilde B_{f,2}=\frac{I_f\left(m^2_2\right)}{F\left(m^2_2\right)}.
\end{equation}
The corresponding partial widths are
\begin{equation}
\label{new_16}
\tilde\Gamma_{f,2}=\tilde B_{f,2}\Gamma_2. 
\end{equation}
Eq.~(\ref{new_15}) is the conventional definition employed in current
calculations, except that the amplitudes are evaluated at the
gauge-independent pole mass $m_2$ rather than at the on-shell mass $M$.
Similarly, Eq.~(\ref{new_16}) also involves the gauge-independent width
$\Gamma_2$ rather than $\Gamma$.
Recalling Eq.~(\ref{def_width_2}), Eq.~(\ref{new_16}) can be expressed in the
form
\begin{eqnarray}
\label{new_18}
m_2\tilde\Gamma_{f,2}=-\frac{I_f\left(m^2_2\right)}{F\left(m^2_2\right)}\,
\frac{I\left(m^2_2\right)}
{1+\left[{\rm Im}\,A(\bar s)-I\left(m^2_2\right)\right]/(m_2\Gamma_2)},
\end{eqnarray}
since the second factor on the r.h.s.\ equals $-m_2\Gamma_2$ \cite{pal}.
We note that the denominator of this second factor differs from the
conventional wave-function renormalization in NNLO. 
By construction, Eqs.~(\ref{new_15}) and (\ref{new_16}) satisfy the additivity
property, $\sum_f\tilde\Gamma_{f,2}=\Gamma_2$. 
In order to establish contact with experiment, 
we now show that it leads to a peak cross section that is gauge independent
through $O(g^4)$, {\it i.e.}\ through NNLO.

Using Eqs.~(\ref{new_1}) and (\ref{new_2}), the amplitude at $s=m_2^2$ for the
process $i\to Z^0\to f$ is found to be
\begin{equation}
\label{new_19}
{\cal A}_{fi}\left(m_2^2\right)
=-\frac{Q_{\mu\nu}V_f^\mu\left(m_2^2\right)V_i^\nu\left(m_2^2\right)}
{m_2\Gamma_2-i\left[A(\bar s)-A\left(m_2^2\right)\right]}
+\tilde N,
\end{equation}
where $\tilde N$ represents non-resonant contributions.
Disregarding for the moment the contributions from $\tilde N$, we consider the
square of the absolute value of the first term on the r.h.s.\ of
Eq.~(\ref{new_19}),
integrate over the phase space of the final-state particles, sum over their
spins, and average over those of the initial-state particles.
Noting that
${\rm Re}\,\left[A(\bar s)-A\left(m_2^2\right)\right]
=m_2\Gamma_2I^\prime\left(m_2^2\right)=O(g^4)$ in leading order, and making
use of Eqs.~(\ref{new_9}), (\ref{new_11}), and (\ref{new_18}), we find for the
resonant contribution through $O(g^4)$ 
\begin{equation}
\label{new_21}
\sigma_{\rm R}^0=\frac{12\pi\tilde\Gamma_{e,2}\tilde\Gamma_{f,2}}
{m_2^2\Gamma^2_2}
\left[1-(I^\prime)^2+\frac{2G\left(m_2^2\right)}{m_2\Gamma_2}\right],
\end{equation}
where we have identified the initial state with an $e^+e^-$ pair and 
$\tilde\Gamma_{e,2}$ ($\tilde\Gamma_{f,2}$) is the $Z^0\to e^+e^-$
($Z^0 \to f$) partial width, defined according to
Eqs.~(\ref{new_15}) and (\ref{new_16}).
In Eq.~(\ref{new_21}), it is understood that $\sigma^0$ is the cross section
devoid of initial-state radiation effects, which are usually taken into
account by a suitable convolution with a {\it Radiator Function} \cite{pdg}. 

At this stage, we recall that the non-resonant amplitude $\tilde N$ in
Eq.~(\ref{new_19}) includes contributions from box and photon-mediated
diagrams as well as non-resonant effects from $\gamma$--$Z$ mixing graphs.
In leading order, their gauge-dependent parts can be found from the results of
Ref.~\cite{deg}.
Inserting Eq.~(\ref{new_14}) into Eq.~(\ref{new_21}), one finds that the
gauge-dependent $(I^\prime)^2$ term is removed and that the ${\rm Im}\,\eta_W$
contribution in Eq.~(\ref{new_14}) cancels in leading order the
gauge-dependent part of the interference of $\tilde N$ with the first term in
Eq.~(\ref{new_19}).
Thus, we find at $s=m^2_2$ 
\begin{equation}
\label{new_25}
\sigma^0\left(m^2_2\right)
=\frac{12\pi\tilde\Gamma_{e,2}\tilde\Gamma_{f,2}}{m^2_2\Gamma^2_2}
\left(1-\frac{\Gamma^2_2}{m^2_2}\right)+\sigma_{\rm B}^0\left(m^2_2\right),
\end{equation}
where we have used $F^\prime\left(m^2_2\right)=-\Gamma_2/m_2$ in leading 
order and the background part $\sigma_{\rm B}^0$ is a gauge-independent
contribution of $O(g^4)$.
As $\sigma^0$ is a physical observable, the fact that Eq.~(\ref{new_25}) is
devoid of gauge-dependent contributions implies that the partial widths 
$\tilde\Gamma_{e,2}$ and $\tilde\Gamma_{f,2}$, defined on the basis of
Eqs.~(\ref{new_15}) and (\ref{new_16}) are gauge independent through $O(g^6)$,
{\it i.e.}\ through NNLO.

The current analyses of the electroweak data measure
$m_1=\left(m^2_2+\Gamma_2^2\right)^{1/2}$ and $\Gamma_1=m_1\Gamma_2/m_2$
rather than $m_2$ and $\Gamma_2$ \cite{si_1}, and determine the peak cross
section at $s=m_1^2$ rather than at $s=m^2_2$.
If the branching ratios are defined by
$\tilde B_{f,1}=I_f\left(m^2_1\right)/F\left(m^2_1\right)$ and the partial
widths by $\tilde\Gamma_{f,1}=\tilde B_{f,1}\Gamma_1 $, instead of
Eqs.~(\ref{new_15}) and (\ref{new_16}), we find that the cross section at
$s=m_1^2$ is given by
\begin{equation}
\label{new_26}
\sigma^0\left(m^2_1\right)
=\frac{12\pi\tilde\Gamma_{e,1}\tilde\Gamma_{f,1}}{m^2_1\Gamma^2_1}
+\sigma_{\rm B}^0\left(m^2_1\right). 
\end{equation}
The theoretical expression employed by the LEP Electroweak Working Group
(EWWG) \cite{ewwg} is of the same form as the first term in
Eq.~(\ref{new_26}).
Thus, Eq.~(\ref{new_26}) justifies this expression through terms of $O(g^4)$,
{\it i.e.}\ through NNLO in the electroweak interactions, 
provided that the gauge-independent definitions of
mass and width are employed!
We note that this means that contributions of $O(g^6)$ to the partial and
total widths are incorporated into Eqs.~(\ref{new_25}) and (\ref{new_26}).
On the other hand, $\sigma_{\rm B}^0$ can be evaluated from tree-level and
one-loop diagrams.

We conclude with the following observations and summary of our results:
(i) In the hadronic sector, the formulation of this paper is restricted to
the parton level of quarks and gluons, {\it i.e.} the effects of confinement
are not taken into account.
Since $\Gamma_\tau$ is of order $10^{-3}$~eV and the widths of the least
massive $B$ and $D$ mesons are even smaller, we neglect the leptonic widths as
well as those of the five lightest quarks.
(ii) If the final-state particles  have negligible widths, Eq.~(\ref{new_4})
provides a gauge-independent definition of partial widths to all orders of
perturbation theory.
In Eqs.~(\ref{new_5}) and (\ref{new_6}), we have shown how this definition
can be modified in order to satisfy the additivity property.
(iii) A rigorous analysis does not include unstable particles in the final
states.
They are rather treated as virtual particles, which decay into stable ones. 
Examples are $Z^0\to f_1\bar f_2W^*\to f_1\bar f_2f_3\bar f_4$,
$H\to W^{+*}W^{-*}\to f_5\bar f_6 f_7 \bar f_8$, where $f_i$ denote stable
fermions.
In domains of phase space where the virtual $W^*$ bosons are in their resonant
regions, a resummation analogous to Eq.~(\ref{new_3}) is in general required. 
In the $W$-boson case, processes of this kind are forbidden by kinematic
considerations.
(iv) In Eq.~(\ref{new_13}), we have analyzed, in leading order and in the
$Z^0$-boson case, the difference $X-m_2\Gamma_2$ between two different
gauge-independent definitions of total widths, based on the pole residues and
the pole position, respectively.
An interesting byproduct is the evaluation of the amplitude
$G\left(m^2_2\right)$ [Eq.~(\ref{new_14})], which represents the contribution
to $I\left(m^2_2\right)$ from unphysical intermediate states.
The result $G\left(m^2_2\right)\neq0$ in $O(g^6)$ reflects the fact that the
unstable particle is not an asymptotic state.
(v) In Eqs.~(\ref{new_15})--(\ref{new_26}), we have examined, in the
$Z^0$-boson case, an alternative and simpler definition of partial widths that
is similar to the one employed in current calculations, except that it
makes use of the {\it pole} rather than the {\it on-shell} mass and width.
Subject to this modification, Eq.~(\ref{new_26}) provides a theoretical
justification, through NNLO in the electroweak interactions, 
for the peak cross section employed by the EWWG \cite{ewwg}.
In this regard, it is important to note that Eqs.~(\ref{new_25}) and
(\ref{new_26}) incorporate corrections of $O(g^6)$ to the width and partial
widths. 

A.S. would like to thank the members of the $2^{\rm nd}$ Institute for
Theoretical Physics of Hamburg University and the Brookhaven National
Laboratory for their warm hospitality during summer 2000, and the Alexander
von Humboldt Foundation for its kind support.  
The research of P.A.G. and A.S. was supported in part by NSF Grants No.\
PHY-9722083 and No.\ PHY-0070787. 
The research of B.A.K. was supported in part by DFG Grant No.\ KN~365/1-1,
by BMBF Grant No.\ 05~HT9GUA~3, and by the European Commission through TMR
Network No.\ ERBFMRX-CT98-0194.

\end{document}